\documentclass[aps,prd,twocolumn,superscriptaddress,showpacs]{revtex4}

\usepackage{graphicx}
\usepackage{dcolumn}
\usepackage{bm}
\usepackage{pstricks}
\usepackage{rotating}
\usepackage{epsfig}
\RequirePackage{xspace}

\def\cdf2{CDF II}
\def\run2{Run II}

\def\c     {\ensuremath{c}\xspace}

\def\b     {\ensuremath{b}\xspace}

\def\bbbar {\ensuremath{b\overline b}\xspace}

\def\pip   {\ensuremath{\pi^+}\xspace}
\def\pim   {\ensuremath{\pi^-}\xspace}

\def\pipm  {\ensuremath{\pi^\pm}\xspace}

\def\Kbar  {\kern 0.2em\overline{\kern -0.2em K}{}\xspace}

\def\Kz    {\ensuremath{K^0}\xspace}
\def\Kzb   {\ensuremath{\Kbar^0}\xspace}
\def\KzKzb {\ensuremath{\Kz \kern -0.16em \Kzb}\xspace}
\def\Kp    {\ensuremath{K^+}\xspace}
\def\Km    {\ensuremath{K^-}\xspace}

\def\KpKm  {\ensuremath{\Kp \kern -0.16em \Km}\xspace}

\def\Dbar    {\kern 0.2em\overline{\kern -0.2em D}{}\xspace}

\def\Dz      {\ensuremath{D^0}\xspace}
\def\Dzb     {\ensuremath{\Dbar^0}\xspace}
\def\DzDzb   {\ensuremath{\Dz {\kern -0.16em \Dzb}}\xspace}
\def\Dp      {\ensuremath{D^+}\xspace}
\def\Dm      {\ensuremath{D^-}\xspace}

\def\DpDm    {\ensuremath{\Dp {\kern -0.16em \Dm}}\xspace}

\def\Bbar    {\kern 0.18em\overline{\kern -0.18em B}{}\xspace}

\def\Bz      {\ensuremath{B^0}\xspace}
\def\Bzb     {\ensuremath{\Bbar^0}\xspace}
\def\BzBzb   {\ensuremath{\Bz {\kern -0.16em \Bzb}}\xspace}
\def\Bu      {\ensuremath{B^+}\xspace}
\def\Bub     {\ensuremath{B^-}\xspace}

\def\BpBm    {\ensuremath{\Bu {\kern -0.16em \Bub}}\xspace}

\def\BorBbar    {\kern 0.18em\optbar{\kern -0.18em B}{}\xspace}
\def\DorDbar    {\kern 0.18em\optbar{\kern -0.18em D}{}\xspace}
\def\KorKbar    {\kern 0.18em\optbar{\kern -0.18em K}{}\xspace}

\mathchardef\Upsilon="7107

\def\proton      {\ensuremath{p}\xspace}

\def\neutron     {\ensuremath{n}\xspace}

\def\ppbar {\ensuremath{p\overline p}\xspace}
\mathchardef\Deltares="7101
\mathchardef\Xi="7104
\mathchardef\Lambda="7103
\mathchardef\Sigma="7106
\mathchardef\Omega="710A
\def\Deltabar{\kern 0.25em\overline{\kern -0.25em \Deltares}{}\xspace}
\def\Lbar{\kern 0.2em\overline{\kern -0.2em\Lambda\kern 0.05em}\kern-0.05em{}\xspace}
\def\Sigbar{\kern 0.2em\overline{\kern -0.2em \Sigma}{}\xspace}
\def\Xibar{\kern 0.2em\overline{\kern -0.2em \Xi}{}\xspace}
\def\Obar{\kern 0.2em\overline{\kern -0.2em \Omega}{}\xspace}
\def\Nbar{\kern 0.2em\overline{\kern -0.2em N}{}\xspace}
\def\Xb{\kern 0.2em\overline{\kern -0.2em X}{}\xspace}

\def\Th      {\ensuremath{\Theta^{+}}\xspace}

\def\xmm     {\ensuremath{\Phi^{--}}\xspace}
\def\xmn     {\ensuremath{\Phi^{0}}\xspace}
\def\phipq   {\ensuremath{\Phi^{0,--}}\xspace}
\def\Phipq   {\ensuremath{\Phi}\xspace}

\def\xstn    {\ensuremath{\Xi(1530)}\xspace}

\def\xm      {\ensuremath{\Xi^{-}}\xspace}
\def\ppXmm   {\ensuremath{\xmm \to \xm \pim}\xspace}
\def\ppXmn   {\ensuremath{\xmn \to \xm \pip}\xspace}
\def\ppphipq {\ensuremath{\phipq \to \xm \pi^{+,-}}\xspace}

\def\Lamb    {\ensuremath{\Lambda }\xspace}

\def\pKs  {\ensuremath{\proton K^0_S }\xspace}
\def\nKp  {\ensuremath{\neutron K^{+} }\xspace}
\def\ximpim  {\ensuremath{\xm \pim}\xspace}
\def\ximpip  {\ensuremath{\xm \pip}\xspace}

\def\Stwopq  {\ensuremath{S=-2}\xspace}

\def\BR         {{\ensuremath{\cal B}\xspace}}

\def\pt         {\mbox{$p_T$}\xspace}

\newcommand{\tev}{\ensuremath{\mathrm{\,Te\kern -0.1em V}}\xspace}
\newcommand{\gev}{\ensuremath{\mathrm{\,Ge\kern -0.1em V}}\xspace}
\newcommand{\mev}{\ensuremath{\mathrm{\,Me\kern -0.1em V}}\xspace}
\newcommand{\kev}{\ensuremath{\mathrm{\,ke\kern -0.1em V}}\xspace}
\newcommand{\ev}{\ensuremath{\mathrm{\,e\kern -0.1em V}}\xspace}
\newcommand{\gevc}{\ensuremath{{\mathrm{\,Ge\kern -0.1em V\!/}c}}\xspace}
\newcommand{\mevc}{\ensuremath{{\mathrm{\,Me\kern -0.1em V\!/}c}}\xspace}
\newcommand{\gevcc}{\ensuremath{{\mathrm{\,Ge\kern -0.1em V\!/}c^2}}\xspace}
\newcommand{\mevcc}{\ensuremath{{\mathrm{\,Me\kern -0.1em V\!/}c^2}}\xspace}

\def\mtr    {\ensuremath{\rm \,m}\xspace}
\def\cm     {\ensuremath{\rm \,cm}\xspace}

\def\mm     {\ensuremath{\rm \,mm}\xspace}

\def\mum    {\ensuremath{\,\mu\rm m}\xspace}

\def\invpb    {\ensuremath{\mbox{\,pb}^{-1}}\xspace}

\def\Tesla {\ensuremath{{\rm\,T}}\xspace}

\def\degrees{\ensuremath{^{\circ}}\xspace}

\def\jetset74   {\mbox{\tt Jetset \hspace{-0.5em}7.\hspace{-0.2em}4}\xspace}

\newcommand{\jprlBase}       {Phys.\ Rev.\ Lett.\xspace}
\newcommand{\jprBase}        {Phys.\ Rev.\xspace}
\newcommand{\jplBase}        {Phys.\ Lett.\xspace}
\newcommand{\nimBaseA}       {Nucl.\ Instr.\ Meth.\xspace}

\newcommand{\zpBase}         {Z.\ Phys.\xspace}

\newcommand{\itns}      [1]  {{IEEE Trans.\ Nucl.\ Sci.\ {\bf #1}}}

\newcommand{\nima}      [1]  {\nimBaseA~A~{\bf #1}}

\newcommand{\npps}      [1]  {{Nucl.\ Phys.\ Proc.\ Suppl.\ {\bf #1}}}

\newcommand{\pan}       [1]  {{Phys.\ Atom.\ Nuclei~{\bf #1}}}

\newcommand{\plb}       [1]  {\jplBase\ B~{\bf #1}}
\newcommand{\prep}      [1]  {{Phys.\ Rep.\ {\bf #1}}}
\newcommand{\jprl}      [1]  {\jprlBase\ {\bf #1}}

\newcommand{\jprc}      [1]  {\jprBase\ C~{\bf #1}}
\newcommand{\jprd}      [1]  {\jprBase\ D~{\bf #1}}

\newcommand{\zpa}       [1]  {\zpBase\ A~{\bf #1}}

\newcommand{\hepex}     [1]  {hep-ex/{#1}}

\newcommand{\ra}{\rightarrow}

\begin{document}

\preprint{CDF-PUB-7361}

  \title{Search for Exotic $\mathbf{S=-2}$ Baryons in $\mathbf{\ppbar}$
         Collisions at $\mathbf{\sqrt{s}=1.96\,TeV}$}

\affiliation{Institute of Physics, Academia Sinica, Taipei, Taiwan 11529, Republic of China} 
\affiliation{Argonne National Laboratory, Argonne, Illinois 60439} 
\affiliation{Institut de Fisica d'Altes Energies, Universitat Autonoma de Barcelona, E-08193, Bellaterra (Barcelona), Spain} 
\affiliation{Baylor University, Waco, Texas  76798} 
\affiliation{Istituto Nazionale di Fisica Nucleare, University of Bologna, I-40127 Bologna, Italy} 
\affiliation{Brandeis University, Waltham, Massachusetts 02254} 
\affiliation{University of California, Davis, Davis, California  95616} 
\affiliation{University of California, Los Angeles, Los Angeles, California  90024} 
\affiliation{University of California, San Diego, La Jolla, California  92093} 
\affiliation{University of California, Santa Barbara, Santa Barbara, California 93106} 
\affiliation{Instituto de Fisica de Cantabria, CSIC-University of Cantabria, 39005 Santander, Spain} 
\affiliation{Carnegie Mellon University, Pittsburgh, PA  15213} 
\affiliation{Enrico Fermi Institute, University of Chicago, Chicago, Illinois 60637} 
\affiliation{Comenius University, 842 48 Bratislava, Slovakia; Institute of Experimental Physics, 040 01 Kosice, Slovakia} 
\affiliation{Joint Institute for Nuclear Research, RU-141980 Dubna, Russia} 
\affiliation{Duke University, Durham, North Carolina  27708} 
\affiliation{Fermi National Accelerator Laboratory, Batavia, Illinois 60510} 
\affiliation{University of Florida, Gainesville, Florida  32611} 
\affiliation{Laboratori Nazionali di Frascati, Istituto Nazionale di Fisica Nucleare, I-00044 Frascati, Italy} 
\affiliation{University of Geneva, CH-1211 Geneva 4, Switzerland} 
\affiliation{Glasgow University, Glasgow G12 8QQ, United Kingdom} 
\affiliation{Harvard University, Cambridge, Massachusetts 02138} 
\affiliation{Division of High Energy Physics, Department of Physics, University of Helsinki and Helsinki Institute of Physics, FIN-00014, Helsinki, Finland} 
\affiliation{University of Illinois, Urbana, Illinois 61801} 
\affiliation{The Johns Hopkins University, Baltimore, Maryland 21218} 
\affiliation{Institut f\"{u}r Experimentelle Kernphysik, Universit\"{a}t Karlsruhe, 76128 Karlsruhe, Germany} 
\affiliation{High Energy Accelerator Research Organization (KEK), Tsukuba, Ibaraki 305, Japan} 
\affiliation{Center for High Energy Physics: Kyungpook National University, Taegu 702-701, Korea; Seoul National University, Seoul 151-742, Korea; and SungKyunKwan University, Suwon 440-746, Korea} 
\affiliation{Ernest Orlando Lawrence Berkeley National Laboratory, Berkeley, California 94720} 
\affiliation{University of Liverpool, Liverpool L69 7ZE, United Kingdom} 
\affiliation{University College London, London WC1E 6BT, United Kingdom} 
\affiliation{Centro de Investigaciones Energeticas Medioambientales y Tecnologicas, E-28040 Madrid, Spain} 
\affiliation{Massachusetts Institute of Technology, Cambridge, Massachusetts  02139} 
\affiliation{Institute of Particle Physics: McGill University, Montr\'{e}al, Canada H3A~2T8; and University of Toronto, Toronto, Canada M5S~1A7} 
\affiliation{University of Michigan, Ann Arbor, Michigan 48109} 
\affiliation{Michigan State University, East Lansing, Michigan  48824} 
\affiliation{Institution for Theoretical and Experimental Physics, ITEP, Moscow 117259, Russia} 
\affiliation{University of New Mexico, Albuquerque, New Mexico 87131} 
\affiliation{Northwestern University, Evanston, Illinois  60208} 
\affiliation{The Ohio State University, Columbus, Ohio  43210} 
\affiliation{Okayama University, Okayama 700-8530, Japan} 
\affiliation{Osaka City University, Osaka 588, Japan} 
\affiliation{University of Oxford, Oxford OX1 3RH, United Kingdom} 
\affiliation{University of Padova, Istituto Nazionale di Fisica Nucleare, Sezione di Padova-Trento, I-35131 Padova, Italy} 
\affiliation{LPNHE, Universite Pierre et Marie Curie/IN2P3-CNRS, UMR7585, Paris, F-75252 France} 
\affiliation{University of Pennsylvania, Philadelphia, Pennsylvania 19104} 
\affiliation{Istituto Nazionale di Fisica Nucleare Pisa, Universities of Pisa, Siena and Scuola Normale Superiore, I-56127 Pisa, Italy} 
\affiliation{University of Pittsburgh, Pittsburgh, Pennsylvania 15260} 
\affiliation{Purdue University, West Lafayette, Indiana 47907} 
\affiliation{University of Rochester, Rochester, New York 14627} 
\affiliation{The Rockefeller University, New York, New York 10021} 
\affiliation{Istituto Nazionale di Fisica Nucleare, Sezione di Roma 1, University of Rome ``La Sapienza," I-00185 Roma, Italy} 
\affiliation{Rutgers University, Piscataway, New Jersey 08855} 
\affiliation{Texas A\&M University, College Station, Texas 77843} 
\affiliation{Istituto Nazionale di Fisica Nucleare, University of Trieste/\ Udine, Italy} 
\affiliation{University of Tsukuba, Tsukuba, Ibaraki 305, Japan} 
\affiliation{Tufts University, Medford, Massachusetts 02155} 
\affiliation{Waseda University, Tokyo 169, Japan} 
\affiliation{Wayne State University, Detroit, Michigan  48201} 
\affiliation{University of Wisconsin, Madison, Wisconsin 53706} 
\affiliation{Yale University, New Haven, Connecticut 06520} 
\author{A.~Abulencia}
\affiliation{University of Illinois, Urbana, Illinois 61801}
\author{J.~Adelman}
\affiliation{Enrico Fermi Institute, University of Chicago, Chicago, Illinois 60637}
\author{T.~Affolder}
\affiliation{University of California, Santa Barbara, Santa Barbara, California 93106}
\author{T.~Akimoto}
\affiliation{University of Tsukuba, Tsukuba, Ibaraki 305, Japan}
\author{M.G.~Albrow}
\affiliation{Fermi National Accelerator Laboratory, Batavia, Illinois 60510}
\author{D.~Ambrose}
\affiliation{Fermi National Accelerator Laboratory, Batavia, Illinois 60510}
\author{S.~Amerio}
\affiliation{University of Padova, Istituto Nazionale di Fisica Nucleare, Sezione di Padova-Trento, I-35131 Padova, Italy}
\author{D.~Amidei}
\affiliation{University of Michigan, Ann Arbor, Michigan 48109}
\author{A.~Anastassov}
\affiliation{Rutgers University, Piscataway, New Jersey 08855}
\author{K.~Anikeev}
\affiliation{Fermi National Accelerator Laboratory, Batavia, Illinois 60510}
% ...
\collaboration{CDF Collaboration\footnote{With visitors from $^a$University of Athens, 
$^b$University of Bristol, 
$^c$University Libre de Bruxelles, 
$^d$Cornell University, 
$^e$University of Cyprus, 
$^f$University of Dublin, 
$^g$University of Edinburgh, 
$^h$University of Heidelberg, 
$^i$Universidad Iberoamericana, 
$^j$University of Manchester, 
$^k$Nagasaki Institute of Applied Science, 
$^l$University de Oviedo, 
$^m$University of London, Queen Mary and Westfield College, 
$^n$Texas Tech University, 
$^o$IFIC(CSIC-Universitat de Valencia), 
}}
\noaffiliation





\date{December 28, 2006}

\begin{abstract}
    A search for a manifestly exotic \Stwopq baryon state decaying to
    $\Xi^-\pi^-$, and its neutral partner decaying to $\Xi^-\pi^+$, 
%
    has been performed using 220\invpb 
    of \ppbar collisions at $\sqrt{s}=1.96$\tev collected
    by the Collider Detector at Fermilab.  The \xm trajectories were
    measured in a silicon tracker before their decay, resulting in a
    sample with low background and excellent position resolution. No
    evidence was found for \Stwopq pentaquark candidates in the invariant
    mass range of 1600--2100\mevcc.
    Upper limits on the product of pentaquark production cross section times its 
branching fraction to  $\Xi^-\pi^{+,-}$,  relative to the cross section of the
well established    \xstn resonance,
    are presented for neutral and doubly negative candidates with \pt$>2$\gevc and
    $\left | y \right |<1$ as a function of pentaquark mass. 
At 1862\mevcc, these upper limits 
for
    neutral and doubly negative final states were found to be
    3.2\% and 1.7\% at the 90\% confidence level, respectively.
\end{abstract}


\maketitle

  Searches for hadrons characterized by exotic quantum
  numbers, i.e., quantum numbers that cannot be obtained from minimal
  quark-antiquark or three quark configurations of standard mesons and baryons,
  have taken place since the outset of the quark model.  Until recently
  these searches yielded little evidence for such states. Most notably
  in the 1960's and 1970's there were searches for a 
  $Z^*$-baryon with positive strangeness in bubble chamber experiments
  with incident beams of pions and kaons, which ended with negative
  results.   These earlier searches are summarized in detailed reviews
   by Hey and Kelly \cite{Hey:1982aj} and more recently by Jennings
   and Maltman \cite{Jennings:2003wz}.

  During 2003, however, the situation changed dramatically. A narrow
  resonance, decaying to \nKp, was reported by the LEPS Collaboration
  in their study of exclusive photon-nucleon
  reactions \cite{Nakano:2003qx}. A state decaying to \nKp must have
  positive strangeness, an exotic quantum number for a baryon, and
  therefore cannot exist in a three-quark model.  A pentaquark
  interpretation was employed, suggesting its quark contents to be
  $uudd\bar{s}$. The mass of the resonance was measured to be 
  $(1540\pm10)$\mevcc, in remarkable agreement with the theoretical prediction
  for the \Th state made in the framework of the chiral soliton
  model by Diakonov, Petrov, and Polyakov \cite{diakonov}. 
  Various experiments, using incident beams of real and quasi-real photons, 
  kaons, and neutrons, claimed to confirm the initial reports of  \Th decaying to \nKp and \pKs  \cite{Barmin:2003vv,Stepanyan:2003qr,Kubarovsky:2003fi,Abdel-Bary:2004ts,
  Aleev:2004sa,Barth:2003ja,Asratyan:2003cb,Airapetian:2003ri,Chekanov:2004kn}.


  Another manifestly exotic state, a doubly negative baryon
   decaying to  $\ximpim$ \cite{cc},
  and its neutral partner decaying to
  \ximpip, were reported by the NA49 Collaboration in
  $\sqrt{s}=17.2$\gev $\proton\proton$ collisions at the CERN SPS
  \cite{Alt:2003vb}.  These states have been tentatively labeled 
  \xmm and \xmn respectively, and the mass of the combined \xmm and
  \xmn signals was measured to be \( (1862\pm2)\mevcc\).  These
  resonances were interpreted as the $I_3=-3/2$ and the $I_3=+1/2$
  partners of an isospin quadruplet of five-quark states with quark
  contents $dsds\bar{u}$ and $dsus\bar{d}$, respectively.
  No independent confirmation of these states exists.


  All ostensible pentaquark signals are based on relatively small data
  samples.  Typically 20 -- 100 signal events are seen with
  estimated significances ranging from three to five standard deviations. 
  With the  exception of the ZEUS and HERMES Collaborations, which both have 
  reported observations of \Th in the \pKs mode,
  many searches for strange pentaquark states produced in high-energy,
  high-statistics, fixed target or collider experiments have yielded
  negative results
  \cite{Aubert:2004ps,Longo:2004gd,Bai:2004gk,Abt:2004tz,
  Schael:2004nm,Aubert:2004bm,Abe:2004ws,Abe:2005gy,Stenson:2004yz,Airapetian:2004mi,
  Adamovich:2004dr,Chekanov:2005at,Ageev:2005ud,Hartouni:1988ic,Armstrong:2004gp,Christian:2005kh,Aubert:2005qi}. 
  More recently, a high statistics analysis of the exclusive reaction $\gamma p\ra\bar{K}^0K^+n$ performed by 
  CLAS, which was one of the experiments initially reporting the observation of the \Th, yielded no evidence 
  for this state \cite{Battaglieri:2005er}.


  We report the results of  searches for exotic baryons
  \phipq, decaying to $\Xi^{-}\pi^{+,-}$, in \ppbar
  collisions at $\sqrt{s}=1.96$\tev. The data were recorded with the
  \cdf2 detector at the Fermilab Tevatron and correspond to
  220\invpb of integrated luminosity.

  The \cdf2 detector is described in detail elsewhere
  \cite{detector}. The aspects of the detector most relevant to this
  analysis are the tracking and the three-stage triggering systems. The
  tracking system lies within a superconducting solenoid, which
  provides a uniform axial magnetic field of 1.4\Tesla.  A system of
  double-sided micro-strip silicon sensors (SVX II) \cite{Sill:2000zz}
  is arranged in five concentric cylindrical shells with radii between
  2.5 and 10.6\cm from the beam-line \cite{coordsystem}, and approximately 90\cm in length. Each silicon layer
  contains axial strips to measure track positions in the plane
  transverse to the beam-line.  Stereo measurements of $\pm$1.2\degrees
  and 90\degrees with respect to the beam-line are also available.
  The remainder of the magnetic volume is occupied by a drift chamber,
  known as the central
  outer tracker (COT) \cite{Affolder:2003ep}.  The COT
  has a radial extent from 40 to 137\cm , and extends 3.1\mtr along the
  direction of the beam.  The sense wires are arranged in 8 super
  layers of 12 wires each, which alternate between axial and 
  $\pm$2\degrees stereo measurements to provide tracking in three
  dimensions.

\begin{figure*}[ht]\centering
\epsfig{file=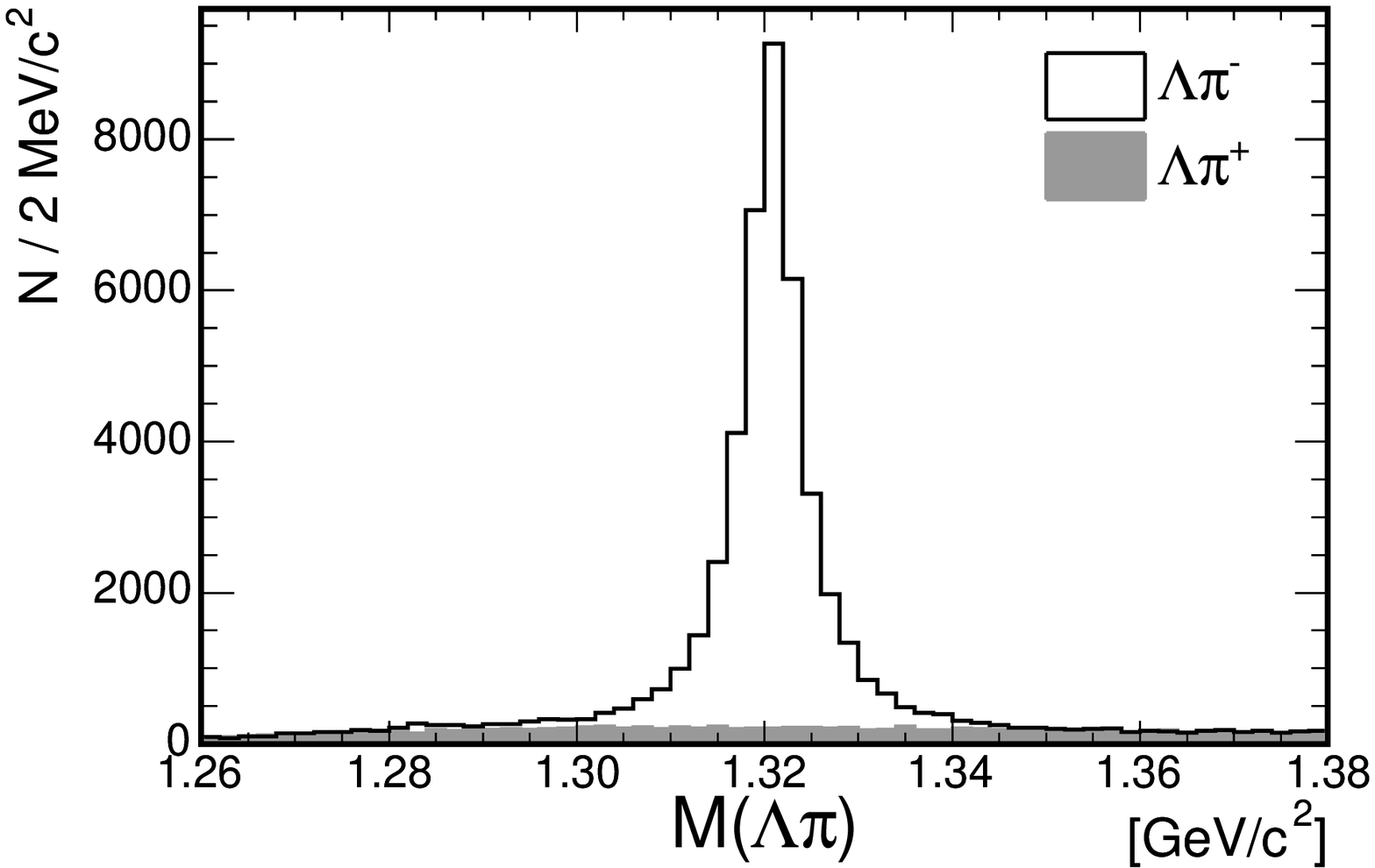,width=52mm,height=40mm}
\epsfig{file=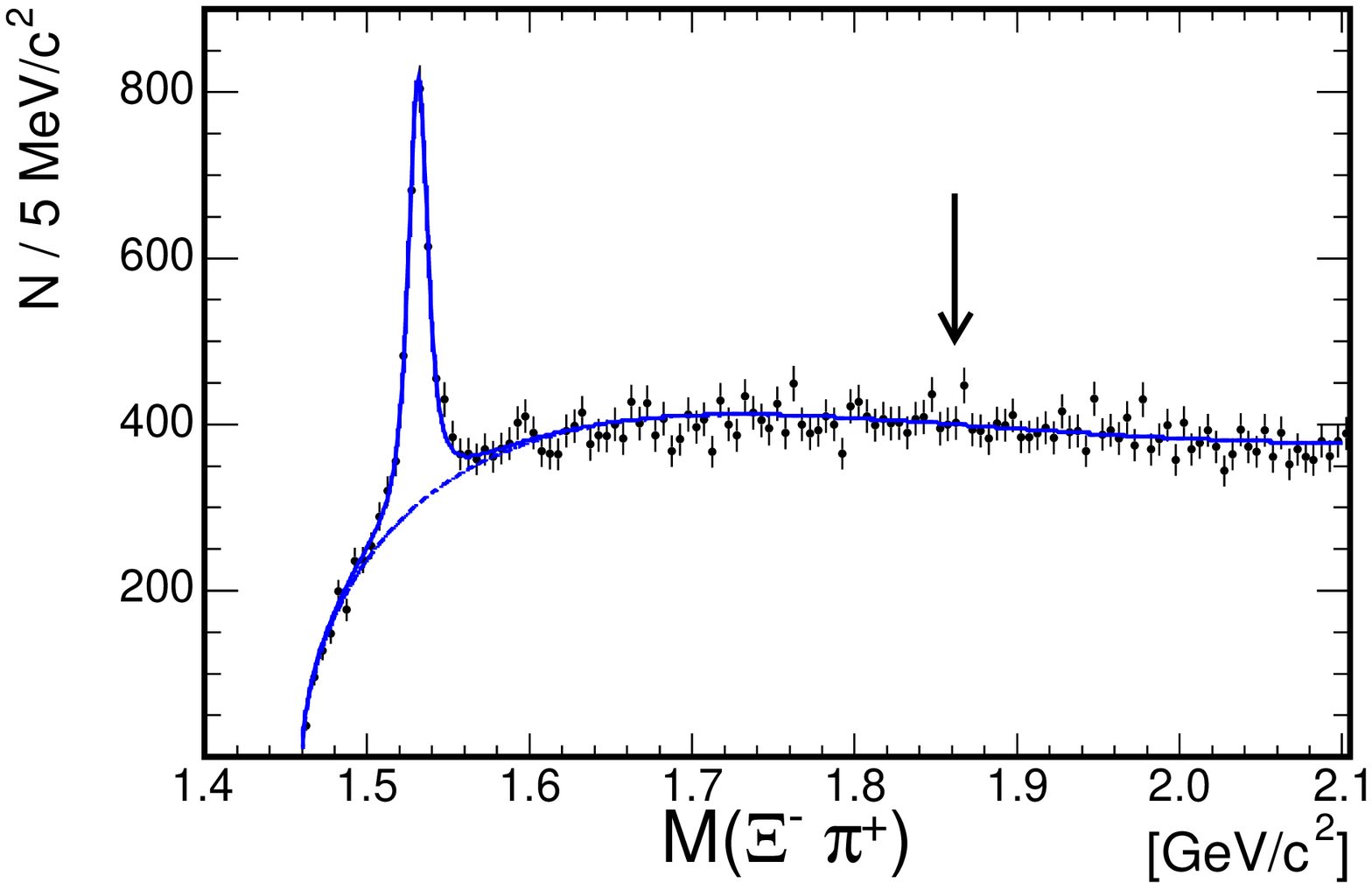,width=52mm,height=40mm}
\epsfig{file=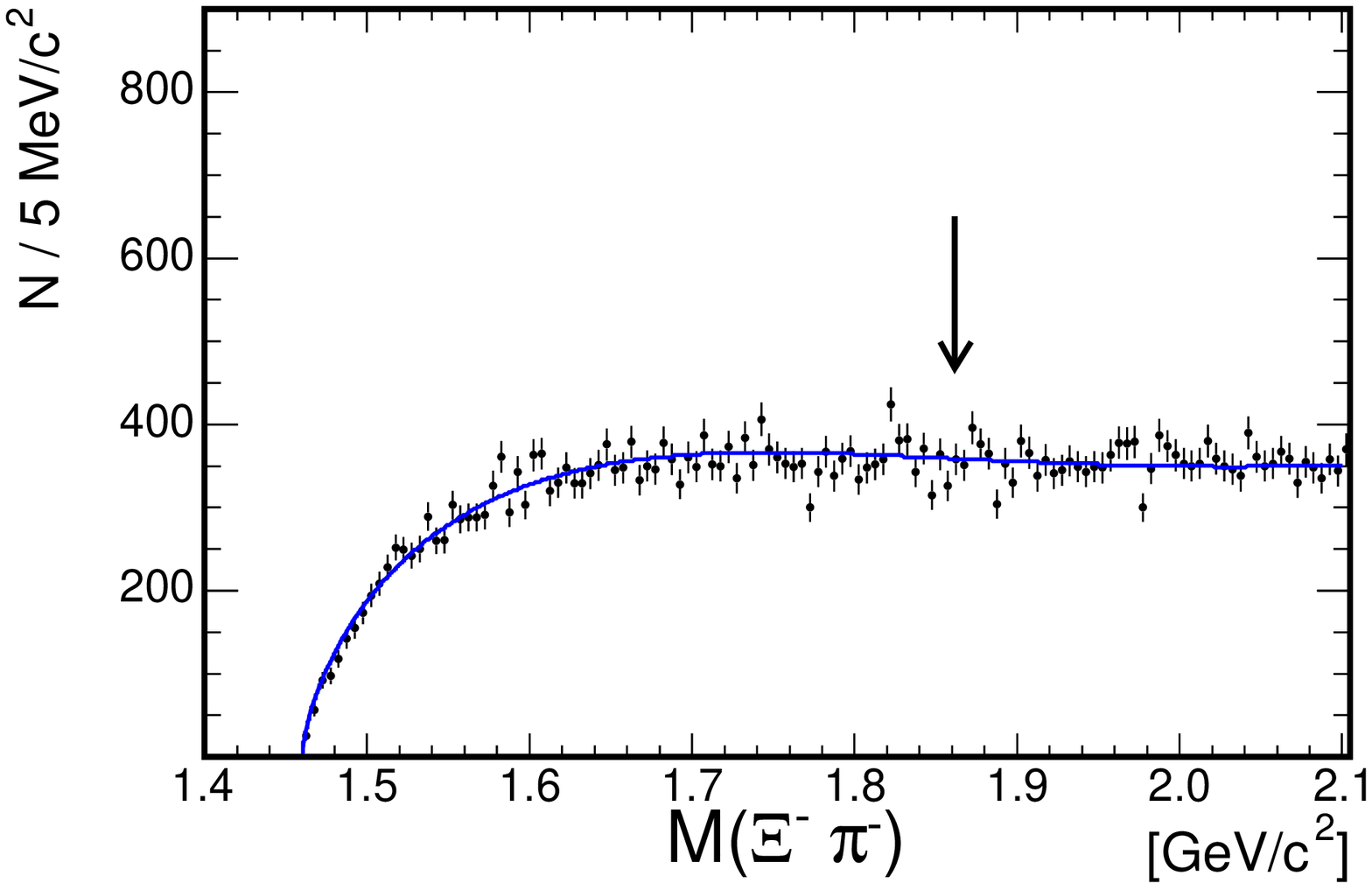,width=52mm,height=40mm}
\rput(-14.75,3){(a)}
\rput(-9.45,3){(b)}
\rput(-4.15,3){(c)}

\caption{Invariant mass spectra from the SVT sample for:
  (a) $\Lambda\pi^\mp$ candidates that have an associated $\Xi^-$
  track in the SVX II, where  the open histogram shows right-sign $\Lambda\pi^-$ combinations,
  and the filled histogram shows wrong-sign $\Lambda\pi^+$ combinations;
  (b) $\Xi^-\pi^+$ combinations, which display a clear $\Xi(1530)$ peak; and
  (c) $\Xi^-\pi^-$ combinations.
  Arrows mark the mass at 1862\mevcc, where the NA49 Collaboration reported observing the $\Phi$.
  The smooth curves represent fits to the spectra and are described in the text.
}

\label{fig:ttt-xi}
\end{figure*}

\begin{figure*}[bth]\centering
\epsfig{file=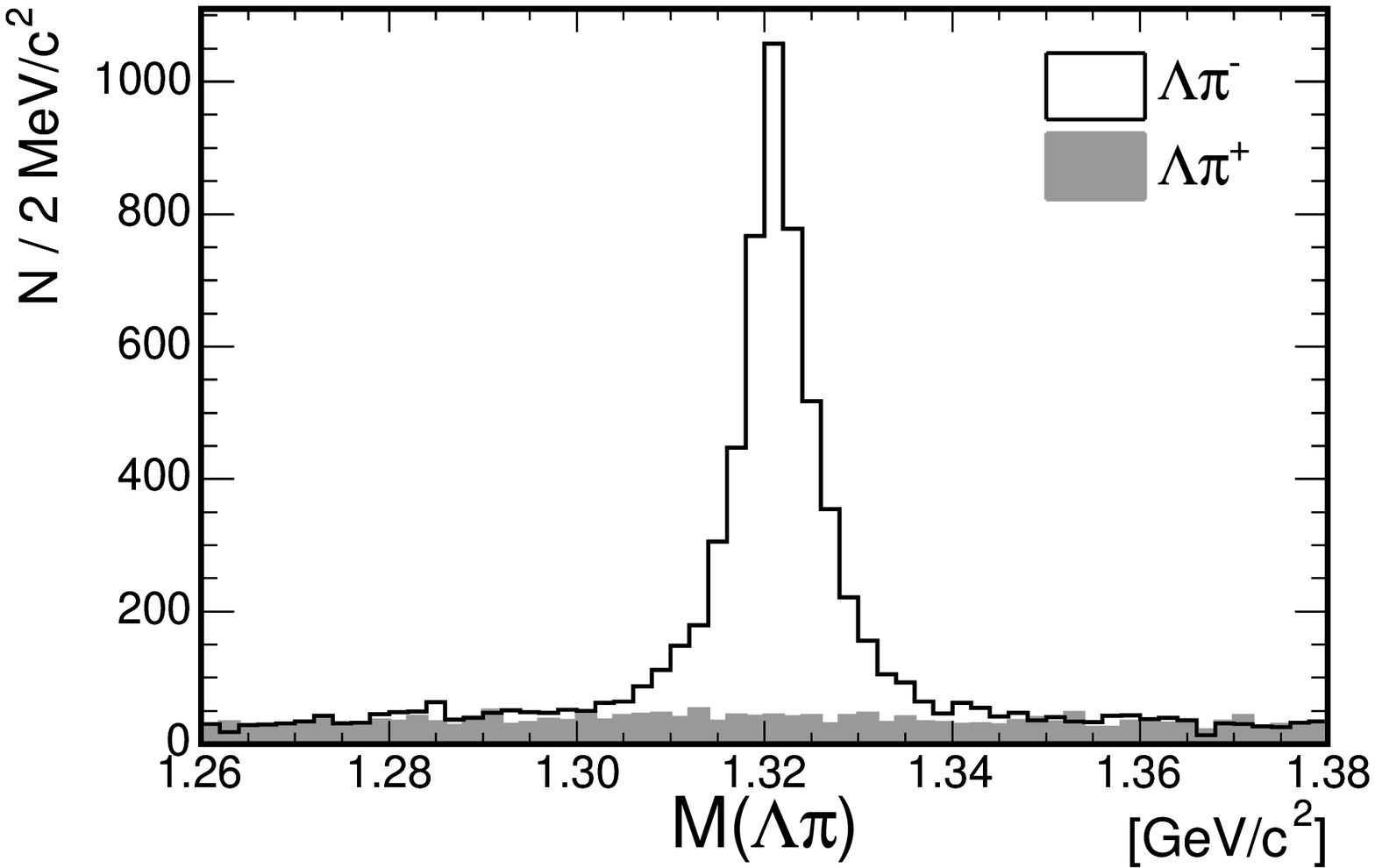,width=52mm,height=40mm}
\epsfig{file=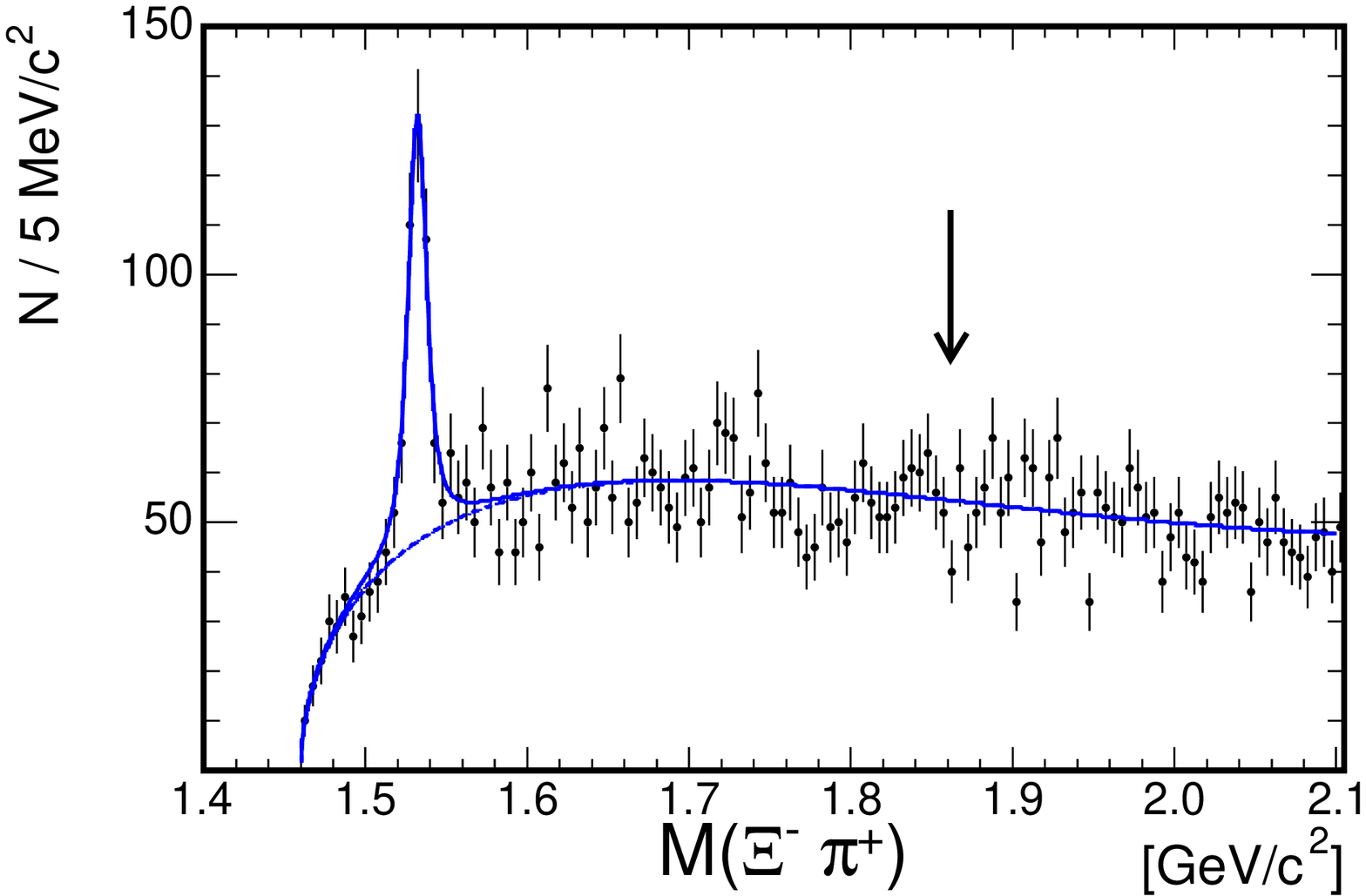,width=52mm,height=40mm}
\epsfig{file=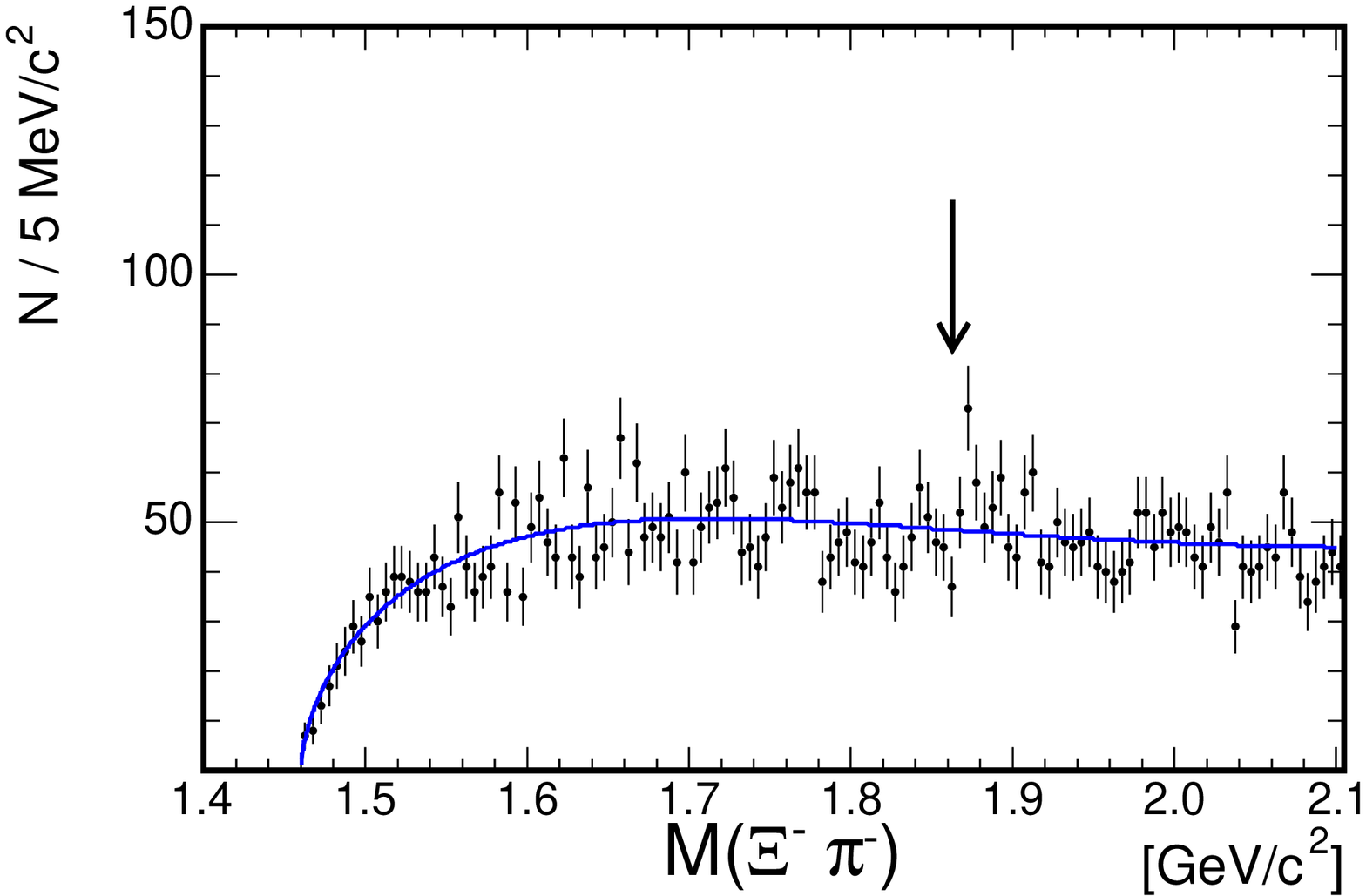,width=52mm,height=40mm}
\rput(-14.75,3){(a)}
\rput(-9.45,3){(b)}
\rput(-4.15,3){(c)}

\caption{
  Invariant mass spectra from the Jet20 sample for:
  (a) $\Lambda\pi^\mp$ candidates that have an associated $\Xi^-$ track, where 
           the open (filled)  histogram is right-sign $\Lambda\pi^-$ 
                (wrong-sign $\Lambda\pi^+$) combinations;
  (b) $\Xi^-\pi^+$ combinations; and
  (c) $\Xi^-\pi^-$ combinations.
  Arrows mark 1862\mevcc, and the smooth curves represent fits described in the text.  
}
%
%
\label{fig:jet20-xipi2}
\end{figure*}

%
The CDF II detector does not have a dedicated trigger selection for $\Phi$ pentaquark
production.
We therefore chose data sets
obtained by two complementary types of triggers.
The first dataset consists of the largest number of inelastic \ppbar collisions
recorded by the experiment, and is based on a displaced-track trigger,
even though the $\Phi$ decay products 
usually produce tracks that have either too small or too large a displacement relative 
to the primary vertex to 
satisfy the trigger requirements. The second sample is obtained 
by a calorimeter-based jet trigger. This trigger has
the merit of being largely independent of the pentaquark production and decay 
properties relevant to the displaced-track trigger.  However, due to its low jet-energy
threshold, only a relatively small fraction
of such triggers could be recorded, thereby limiting
its statistical sensitivity for this search. 

The displaced-track trigger was designed primarily for recording
hadronic B-decays. This trigger requires the presence in the event of a pair 
of oppositely charged tracks identified 
in the transverse view by 
a fast hardware track 
fitter called the silicon vertex trigger (SVT) 
\cite{xft}.
Each track is required to have transverse momentum \(\pt>2.0\gevc\), and  an  impact parameter ($d_0$) which falls
 within the range 
  \({\rm 120\mum} < \left | d_0 \right | < {\rm 1\mm}\) \cite{ip}.
In addition,   the intersection point of the triggered 
track pair is required to have a 
  200\mum\ transverse displacement with respect to the beam-line, 
the scalar sum of the two transverse momenta must exceed
 5.5\gevc, and the azimuthal angular difference between the
 tracks is required to be within the range $2^\circ<\Delta\phi<90^\circ$.

 The calorimeter based trigger 
  requires the presence of a jet with transverse energy
  $E_T>20$\gev (the Jet20 trigger).  The \cdf2 calorimeter
  system covers a pseudorapidity region of 
  \(\left | \eta \right | <3.5 \) \cite{coordsystem}
and consists 
  of electromagnetic and hadronic 
  sections 
\cite{Balka:1987ty}. 
  The physical towers are grouped into trigger towers 
that span $15^\circ$ in
  $\phi$ and approximately 0.2 in $\eta$ for central towers ($|\eta | < 1.0$).   
  The level-1 trigger requires
  a single tower with an $E_T>5\gev$. At level 2, a
  calorimeter cluster is required to have an $E_T$ above
  15\gev. The level-3 trigger uses a jet cone algorithm to
  reconstruct jets with a cone size 
  \(R = \sqrt{\Delta\eta^2+\Delta\phi^2}<0.7\) 
  around a seed tower 
\cite{Arnison:1983rn}.
  Due to trigger rate limitations, the jet trigger is prescaled such that 
   this dataset
  contains an effective integrated luminosity of only about
  0.52\invpb.

  Pentaquark candidates were reconstructed through the decay chain:
  \ppphipq, \(\xm\to\Lamb\pim\) and \(\Lamb\to\proton\pim\).  
  Our reconstruction and selection strategy 
  was guided by using the \(\xstn\to\Xi^-\pi^+\) signal
  in our data as a model for $\Phi$ production.
  The \Lamb candidates were
  reconstructed from oppositely charged pairs of tracks 
  that had more
  than 20 axial and 16 stereo COT hits and 
  an impact parameter 
  greater
  than 500\mum. The track with the highest transverse
  momentum in the pair was assigned the proton mass.  This assignment
  is always correct for true \Lamb candidates  whose 
decay products have sufficient $p_T$ to be measured by the tracking system.
  A vertex constrained fit of track pairs was performed for the
  \Lamb candidates. The invariant mass was required to be 
   within $\pm$5\mevcc of the world average \Lamb mass, and the
  vertex displacement in the transverse plane was required to be greater than 1\cm from the beam-line.

  The \xm hyperons were reconstructed in the decay chain
  \(\xm\to\Lamb\pim\) by combining \Lamb candidates with
  each remaining negative track, assumed to be pion,  in the event.  The track parameters
  of each three-track combination were refit with a \xm decay
  hypothesis, including a mass constraint to the \Lamb mass
  (1115.683\mevcc)~\cite{pdg} on the tracks originating from the \Lamb
  candidate.  Candidates that were consistent with this
  hypothesis, and had a total invariant mass within 60\mevcc of the
  \xm mass (1321.31\mevcc)~\cite{pdg}, were refit with
  a \xm mass constraint. The \xm decay vertex was required
  to be closer to the primary vertex than the \Lamb decay vertex,
  and we required at least 1\cm radial separation between the
  two decay vertices.

  The lifetime of the \xm, \(c\tau=(4.91\pm0.04)\cm\)~\cite{pdg},
  and the \pt requirements resulting from our acceptance, produce a
  \xm decay vertex that lies outside the inner layers of the SVX II 
for a
  majority of the events.  This makes it possible to
  reconstruct the \xm track from the hits left in the layers of
  the SVX II.
  The momentum and decay vertex of the \xm candidate as reconstructed in 
  the COT were used to
  define its track helix. Hits in the SVX II detector made by real
  \xm hyperons are expected to fall near the extrapolation of this
  helix in the silicon detector.  A dedicated tracking algorithm was
  used to attach hits in the SVX II to the \xm helix, which required
  a minimum of two axial silicon hits.  The hyperon tracks found in
  the SVX II were used, with \xm mass assignment, in the subsequent
  analysis in the same way as standard tracks produced by stable particles.


  The $\Lamb\pim$ invariant mass spectrum of \xm candidates
  found in the SVT dataset is shown in Fig. \ref{fig:ttt-xi}(a).  A
  \xm signal containing about 36,000 candidates is evident. The
  invariant mass spectrum of wrong-charge combinations,
  $\Lamb\pip$, shown by the filled histogram, provides a
  good description of the right-sign combinatorial background.  No
  artificial signal was created by the selection criteria or the
  hyperon tracking procedure.

  To search for new states, each \xm candidate was combined with
  all remaining tracks.  These
  tracks were assigned the pion mass and were required to have
  $\pt>0.4$\gevc, at least 20 axial and at least 20 stereo COT hits,
  and 3 or more axial SVX II hits. In addition, \xm tracks and pion
  tracks were required to have impact parameters 
  that were smaller than 0.015\cm and 0.1\cm
  respectively. The remaining $\xm\pipm$ pairs were subjected
  to a common vertex fit with the total momentum of the pair
  constrained to point back to the primary vertex in the transverse plane.
  A vertex fit
  quality requirement was imposed, so that only well measured
  candidates were retained for further analysis.

  No additional kinematic cuts were applied to $\xm\pipm$
  pairs.  
  As described earlier, such
  states are unlikely to produce track pairs satisfying the SVT
  trigger criteria.  
  Therefore,
  $\xm\pipm$ candidates were not required to satisfy the SVT
  trigger. Since the SVT trigger selects events containing particles
  produced in the decays of long-lived \c and \b hadrons, our source
  of pentaquark candidates would be  predominantly the fragmentation of the available
  light quarks in the event, as is the case for \(\xstn\) baryons.


  The invariant mass spectra of \ximpip and \ximpim 
  combinations are shown in Figs. \ref{fig:ttt-xi}(b) and (c) respectively.
  A prominent peak corresponding to the decay $\xstn\to\ximpip$ is
  clearly visible in the $\ximpip$ mass distribution. No other narrow
  structures can be seen in either distribution. The smooth curves
  superimposed on the $\xm\pipm$ spectra show the results of unbinned
  maximum-likelihood fits to the data. The fit function included a
  Breit-Wigner resonance shape convoluted with a Gaussian resolution
  function describing the \xstn, plus a third degree polynomial
  multiplied by an exponential as a model for the background. 
  The width of the Gaussian was allowed
  to float, and the intrinsic width of the \xstn  was
  Gaussian-constrained in the fit to the world average value of $(9.1\pm0.5)\mevcc$
  \cite{pdg}.  The fit yielded $1923\pm80({\rm stat})$ candidates above
  the background at a  mass of $(1531.7\pm0.3(\rm{stat}))$\mevcc, in
  excellent agreement with world average mass of the \xstn.  The
  fitted mass resolution of $(3.5\pm0.6\rm{(stat)})$\mevcc is in
  agreement with the expected mass resolution based on
 the full detector
  simulation.  No excess of events appears at a mass of 1862 \mevcc,
 the mass claimed by NA49 for the $\Phi$ states.
  Since the production mechanism of pentaquark states is unknown, it
  is possible that the SVT trigger bias may preferentially reject
  events with pentaquarks. To address this issue, an identical
  analysis has been performed using the data obtained with the Jet20
  trigger.


  The invariant mass spectra of $\Lamb\pim$ and $\Lamb\pip$
  combinations reconstructed in the Jet20 data and having an associated 
SVX II hyperon 
  track  are shown in Fig. \ref{fig:jet20-xipi2}(a). The jet-trigger
  data sample is significantly smaller than the data sample obtained
  with the trigger on displaced tracks.  However, this sample of 
  $4870\pm128{\rm(stat)}$ \xm candidates is still a factor of 2.5 times larger
  than the sample used by the NA49 experiment in their analysis. The \ximpip 
  and \ximpim invariant mass spectra are shown in Figs.
  \ref{fig:jet20-xipi2}(b) and (c). Again, the baryon \xstn is seen, and no
  narrow structures are visible at the mass of 1862\mevcc. 
  The \ximpip invariant mass spectrum was fit in a 
  similar fashion as previously described with the exception that the
  Gaussian resolution of the \xstn peak was constrained to 
  $(3.5\pm0.6)\mevcc$. The fit yielded a signal of 
  $313\pm\,28{\rm(stat)}$ \xstn candidates.

  There is no indication of \Stwopq pentaquark production in these
  data. We determined an upper limit on the \Phipq
  pentaquark cross section times the branching fraction
  ${\cal B}(\Phipq\ra\Xi^-\pi^{\pm})$
  relative to the \xstn cross section  
  in \ppbar collisions. Knowledge of this ratio
  can be an important input for pentaquark production models.
  Since the two states share the same final state, many sources of
  systematic uncertainty cancel out in this ratio.  We measured this
  ratio for the \Phipq mass of $(1862\pm2)$\mevcc, the value reported by the NA49
  Collaboration, and performed a scan of this ratio in the mass
  range [1600--2100]\mevcc.

\begin{figure}[bth]\centering
\epsfig{file=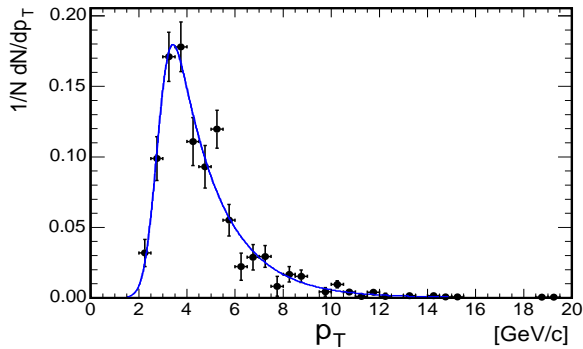,width=80mm,height=50mm}
\caption{Reconstructed transverse momentum spectrum of the \xstn. 
         The fit function is described in the text.}
\label{fig:efficiency}
\end{figure}

  Relative acceptances and reconstruction efficiencies for
  $\Xi^-\pi^{\pm}$ final states of different invariant mass have been
  studied with $\ppbar\to\bbbar\xstn\,X$ events produced
  with the {\sc pythia} event generator~\cite{pythia} and subjected to the
  full {\sc geant} simulation of the CDF II detector, trigger, and
  reconstruction. 
  The pentaquark states were represented in the
  {\sc pythia} decay table as \xstn with various masses. 
  The pentaquarks were not required to originate from hadrons containing a
  \b or \c quark.  
  The simulation of the associated production with \bbbar pairs 
  is a purely practical 
  approach to satisfy SVT trigger conditions.
  Four samples were generated for masses of 1530,
  1696, 1862, and 2027\mevcc in order to obtain the mass dependence of the
  detector resolution and detector efficiency for the whole mass range
  of interest.  
  The generated \pt spectrum of the \xstn was
  re-weighted to reproduce the \pt distribution of the \xstn 
  measured in the SVT trigger data \cite{fit}.
  The latter is shown in Fig.
  \ref{fig:efficiency}. The smooth curve shows a fit with an exponential
  convoluted with a Gaussian. The same weight function was applied
  to the \pt spectra of the generated pentaquarks in all four
  samples.   Lacking any established model for  pentaquark production,
  we assumed
  that the \pt spectrum of the \Phipq pentaquark is the same as that  
  of the \xstn,
  other than the kinematic effects of heavier masses in the {\sc pythia} generation.

%

\begin{figure*}[bth]\centering
\epsfig{file=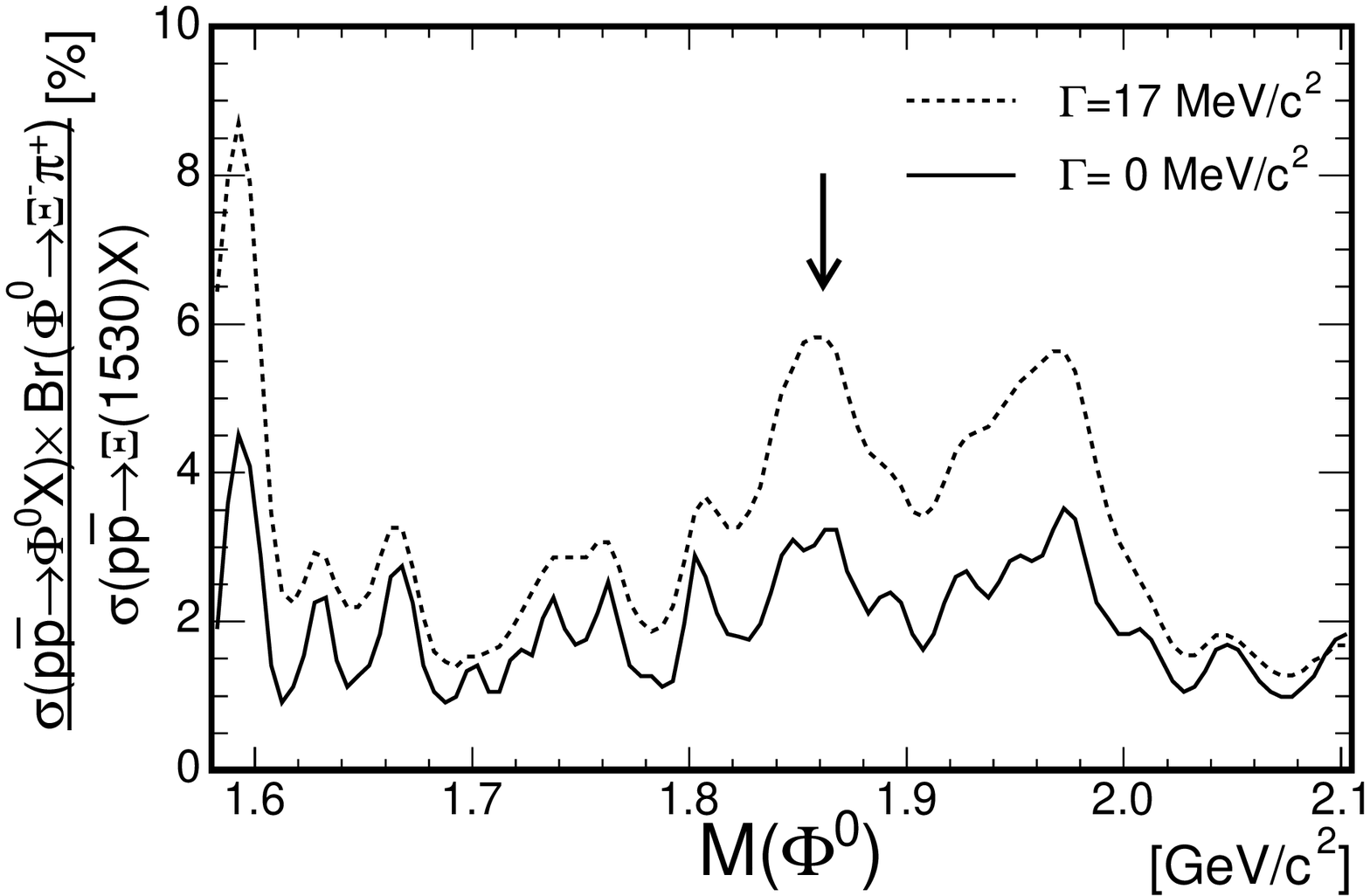,width=80mm,height=50mm}
\epsfig{file=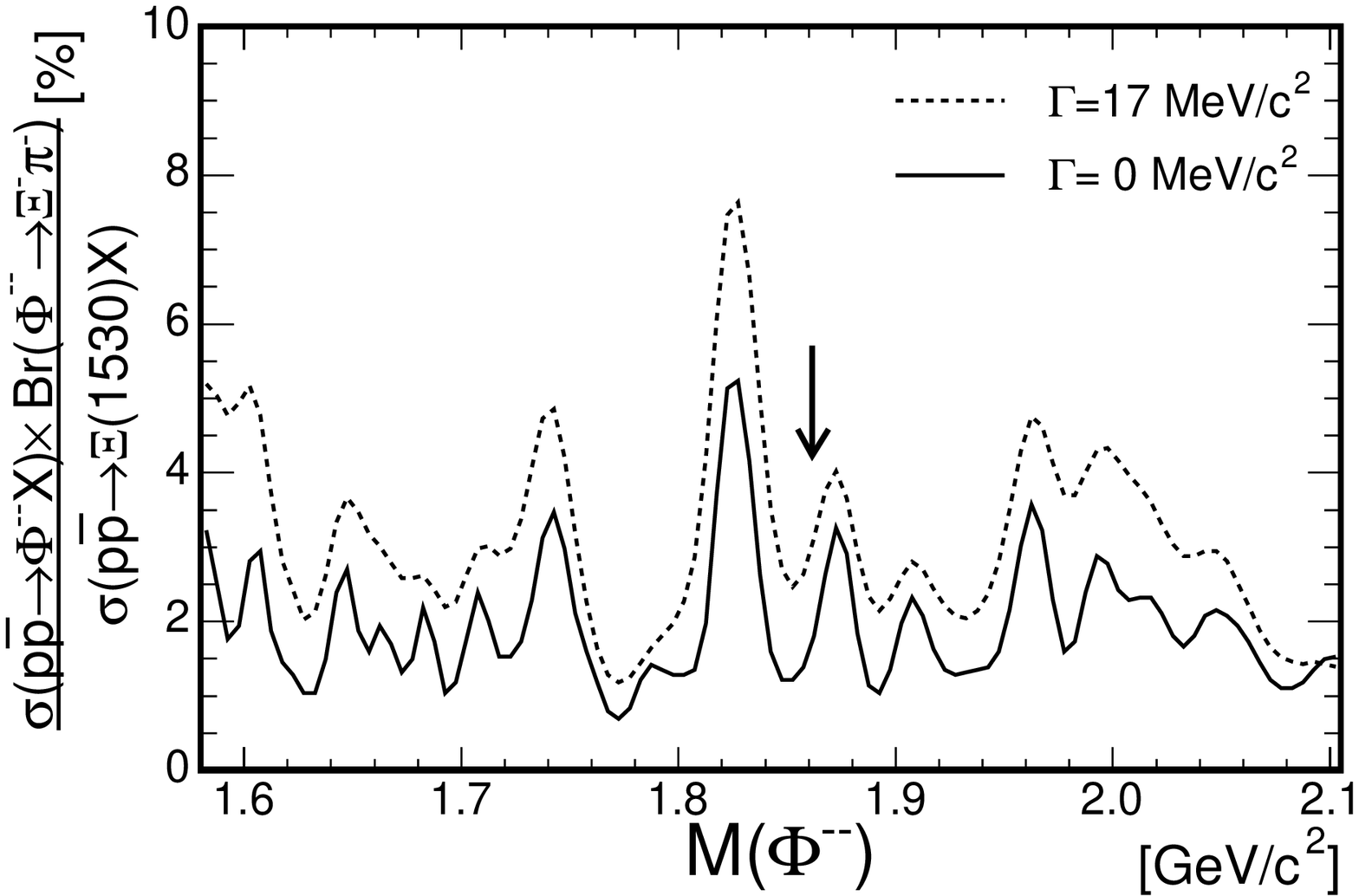,height=50mm}

\rput(-6,4){(a)}
\rput(2,4) {(b)}
\caption{
Upper limits from SVT data (90\% confidence level)
for the production cross section of $\Phi$ pentaquarks times 
the $\Phi\rightarrow \Xi\pi$ branching ratios relative 
to the $\Xi(1530)$ cross section as a function of:
(a) $\Phi^0$ mass, and
(b) $\Phi^{--}$ mass.
The solid line shows the results obtained assuming zero natural width of the pentaquark,
and the dashed line shows the results obtained assuming $\Gamma=17$ \mevcc.
The arrows mark the position of the NA49 peak.
}
\label{fig:ul_ttt}
\end{figure*}

\begin{figure*}[bth]\centering
\epsfig{file=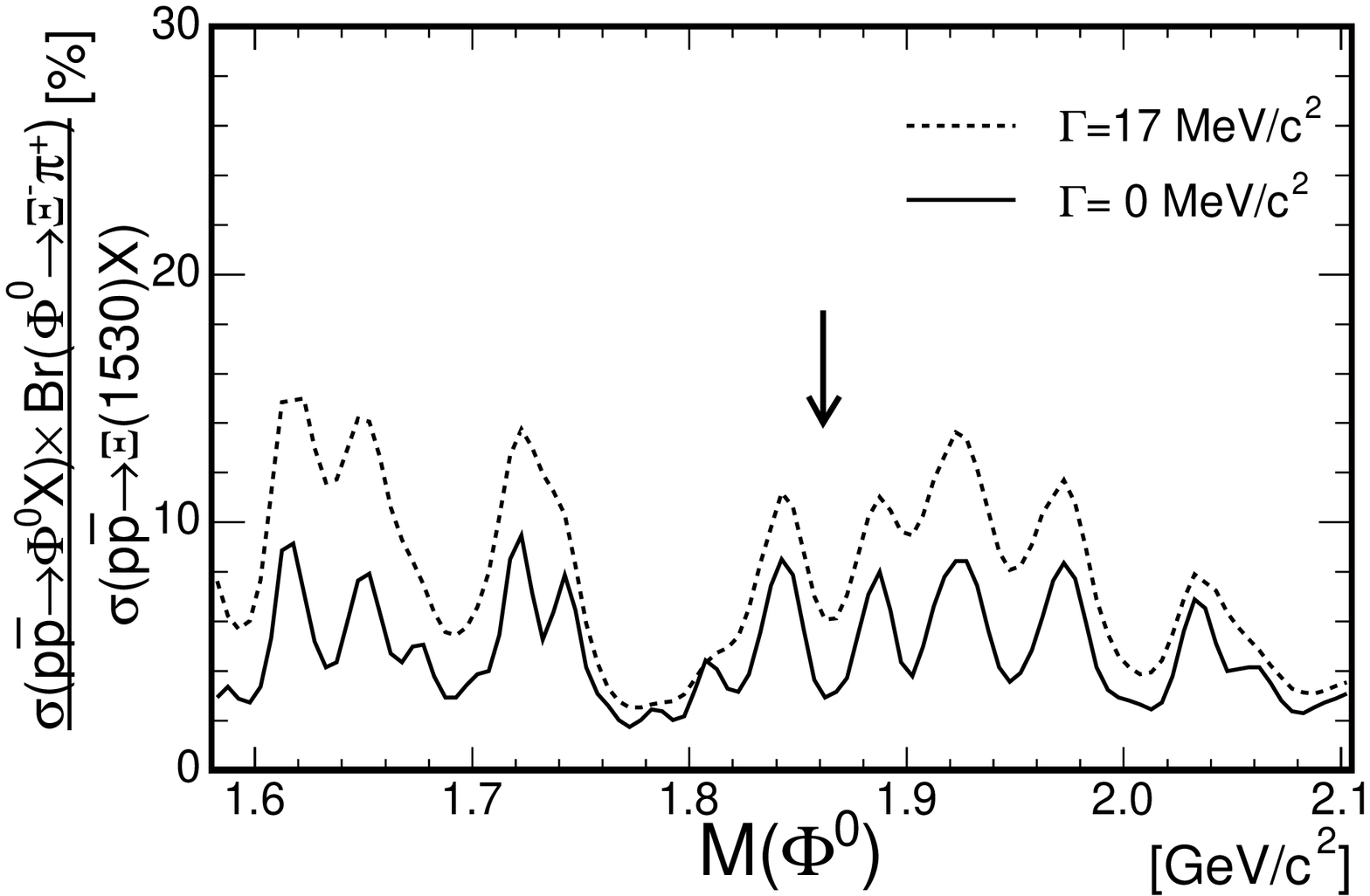,width=80mm,height=50mm}
\epsfig{file=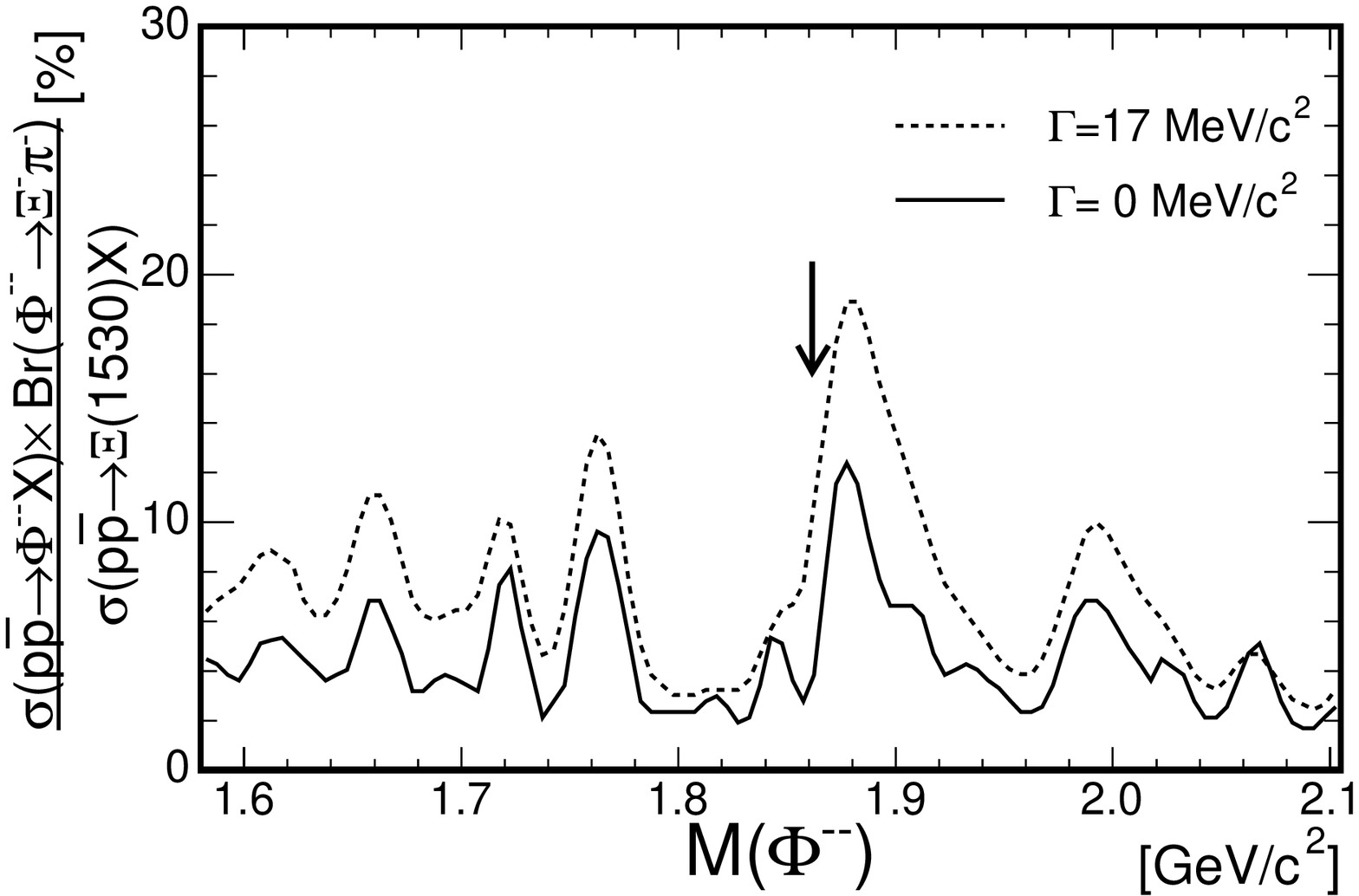,width=80mm,height=50mm}
\rput(-14.5,3.7){(a)}
\rput(-6.4,3.7) {(b)}

\caption{
Upper limits from Jet20 data (90\% confidence level)
for the production cross section of $\Phi$ pentaquarks times 
the $\Phi\rightarrow \Xi\pi$ branching ratios relative to 
to the $\Xi(1530)$ crossection as a function of:
(a) $\Phi^0$ mass, and
(b) $\Phi^{--}$ mass.
The solid (dashed) line shows the results obtained assuming $\Gamma=$ 0 (17) \mevcc.
The arrows mark the position of the NA49 peak.
}
\label{fig:ul_jet20}
\end{figure*}

  \begin{table*}[bth]\centering
    \begin{tabular}{c|c|c}\cline{2-3}
      & SVT trigger & Jet20 trigger \\ 
      \hline 
      ${N(\xm)}$                                & $35722\pm326$                  
                                                & $4870\pm122$ \\ 
      ${N(\xstn)}$                              & $1923\pm 80$                   
                                                & $313\pm28$ \\ 
%
\hline
      \({\displaystyle \frac{\sigma(p\bar{p}\ra{\xmn}X)\cdot \BR(\ppXmn)}
	    {\sigma(p\bar{p}\ra{\xstn}X) }[\%]}\) & ${\rm <3.2~(5.8)}$ 
                                                & ${\rm <3.0~(9.2)}$\\
      \({\displaystyle \frac{ \sigma(p\bar{p}\ra{\xmm}X)\cdot \BR(\ppXmm)}
	     {\sigma(p\bar{p}\ra{\xstn}X)}[\%]}\) & ${\rm <1.7~(3.1)}$ 
                                                & ${\rm <3.2~(10.1)}$\\ 
      \hline
    \end{tabular}
    \caption{Event yields of \xm, \xstn,  
      and upper limits on 
      $\sigma(p\bar{p}\ra\Phi^{0,--} X)\cdot{\cal B}(\Phi^{0,--}\ra\Xi^-\pi^{+,-})/\sigma(p\bar{p}\ra\xstn X)$
      for m(\Phipq) = \((1862\pm2)\mevcc\), 
      at 90\% confidence levels obtained by assuming 
	     zero natural width of the pentaquarks. Numbers in 
	     parentheses were calculated with the assumption of 
	     a 17\mevcc natural width.}

    \label{tab:xi}
  \end{table*}

  The sensitivity of the experiment to $\xm\pipm$ final states is
  limited by our tracking acceptance to particles that have transverse
  momentum greater than 2\gevc\ and $\left|y\right|\leq 1$.
  Compared to the \xstn, a
  pentaquark with a mass of 1862\mevcc would have a factor of 1.5 higher
  detection efficiency at $\pt=2$ GeV/$c$;  this ratio of efficiencies
  decreases with increasing $\pt$.  
We estimate that the ratio of detector
  efficiencies averaged over the range of $\pt$ within our acceptance
  is $1.29\pm0.05\rm{(stat)}$, where the uncertainty
  is due to the size of the simulated event sample. There is an
  additional 10\% systematic uncertainty due to the 
parameterization of the \xstn transverse momentum. 
This value is obtained by varying the parameters of
  the fit to the observed \xstn $p_T$ spectrum within statistical
  errors, and re-weighting the input spectrum of the simulation
  accordingly.

  The limits on $\sigma(p\bar{p}\ra\Phi^{0,--} X)\cdot{\cal B}(\Phi^{0,--}\ra\Xi^-\pi^{+,-})/\sigma(p\bar{p}\ra\xstn X)$ 
  were obtained by introducing an additional signal term
  for the hypothetical pentaquark signal, represented by a
  Breit-Wigner resonance shape convoluted with a Gaussian resolution,
  into the previously described fit function.  The expected detector
  resolution at $m(\xm\pipm)=1862$\mevcc is 
  \( (6.6\pm0.7{\rm(stat)}\pm0.5{\rm(syst))}\mevcc \), 
  which  was obtained through
  detector simulation. 
  The presence of a \Phipq signal was evaluated
  for two hypotheses for its natural width: a natural width of zero,
  and a natural width equal to 17\mevcc, the latter value being the 
  measured full width at half maximum of the peak reported by the NA49 Collaboration
  \cite{Alt:2003vb}.  
  The values of the relative
  efficiency and the mass resolution at $m(\Phipq)=1862$\mevcc  were
  allowed to vary by their uncertainties by adding a Gaussian constraint to the 
  likelihood function.
  The widths of the Gaussians were set to the sum in quadrature  of the
  statistical and systematic uncertainties. We determined the limit
  on the production of the \Phipq observed by the 
NA49 Collaboration by imposing their
  measured mass with a Gaussian constraint of 2\mevcc in the
  likelihood.
  The limits are established using a Bayesian approach, with a 
  uniform prior for the relative production cross section, and 
  taking into account $ \BR(\xstn\to\ximpip)=0.67$ \cite{pdg}. The 
  limits at the 90\% confidence level  are summarized in Table \ref{tab:xi}.

  Available predictions for \Stwopq pentaquark masses vary over a range
  from 1750 to 2027\mevcc
  \cite{diakonov,Jaffe:2003sg,Ellis:2004uz,Karliner:2004uj}. 
  Since we failed to see evidence for a signal at the mass stated by
  the NA49 Collaboration, we have also derived a limit on the relative pentaquark
  production rate as a function of mass in the range
  [1600--2100]\mevcc in order to provide a generic bound on \Phipq
  pentaquark production models.  The procedure to determine the upper
  limit on the relative cross section 
  for a pentaquark mass within the range
  [1600--2100]\mevcc was similar to the one used for testing the NA49
  Collaboration's observation.  We performed a fit for the relative cross section
  over this mass range in 2\mevcc steps in pentaquark mass.  The
  relative detection efficiency and mass resolution entered as a
  function of mass into the fit function. The results of the relative
  cross section scans for neutral and doubly negative pentaquarks are
  shown in Figs. \ref{fig:ul_ttt} and \ref{fig:ul_jet20}.
  Neither the SVT trigger data nor the Jet20
  trigger data show  evidence for pentaquarks over
  this broad mass range.

  In conclusion, we have performed a search for exotic \Stwopq baryon
  states decaying to \ximpim and \ximpip over
  the mass range of [1600--2100]\mevcc. 
 No evidence for a signal has been found, and we have set upper limits on 
the relative cross section times branching fraction of exotic cascade 
baryons with respect to the well established resonance \(\xstn\)  
  for $\pt>2\gevc$ and $\left | y \right |<1$. 
  We summarize our results by quoting the higher statistics
  analysis with the displaced-track trigger, in which  
  $\sigma(p\bar{p}\ra\Phi X)\cdot{\cal B}(\Phi^{0,--}\ra\Xi^-\pi^{+,-})/\sigma(p\bar{p}\ra\xstn X)$
  for neutral and doubly negative states 
  were respectively found to 
  be less than
  3.2\% and 1.7\%  at the 90\% confidence level, for the mass
  reported by the NA49 Collaboration
  \cite{Alt:2003vb}. Consequently, we find that these data are unable
  to confirm the existence of exotic \Stwopq baryons.


We thank the Fermilab staff and the technical staffs of the participating institutions for their vital contributions. This work was supported by the U.S. Department of Energy and National Science Foundation; the Italian Istituto Nazionale di Fisica Nucleare; the Ministry of Education, Culture, Sports, Science and Technology of Japan; the Natural Sciences and Engineering Research Council of Canada; the National Science Council of the Republic of China; the Swiss National Science Foundation; the A.P. Sloan Foundation; the Bundesministerium f\"ur Bildung und Forschung, Germany; the Korean Science and Engineering Foundation and the Korean Research Foundation; the Particle Physics and Astronomy Research Council and the Royal Society, UK; the Institut National de Physique Nucleaire et Physique des Particules/CNRS; the Russian Foundation for Basic Research; the Comisi\'on Interministerial de Ciencia y Tecnolog\'{\i}a, Spain; the European Community's Human Potential Programme under contract HPRN-CT-2002-00292; and the Academy of Finland.

%

\begin{thebibliography}{10000}
\bibitem{cc} Unless otherwise noted, the charge conjugate state is implied.
\bibitem{coordsystem} CDF uses a cylindrical coordinate system 
about the proton beam axis in which $\theta$ is the polar angle, $\phi$ is the azimuthal angle, 
and the pseudorapidity is defined as \( \eta=-\ln[\tan(\theta/2)] \). The true rapidity is 
$y=\frac{1}{2}\ln{\left(\frac{E+p_z}{E-p_z}\right)}$ where 
$E$ is the energy of the particle and $p_z$ is the 
projection of its momentum to $z$-axis.
\bibitem{ip} 
	 The impact  parameter is the distance of closest approach between a track and the primary 
         vertex in the transverse plane.
\bibitem{fit} The spectrum observed in the data is described by the exponential 
            \( dN/d\pt\propto\exp{(-\alpha\pt)} \)
            with \( \alpha=0.70~{({\rm GeV}/c)}^{-1} \).
\end{thebibliography}
\end{document}